\documentclass[reqno,11pt]{amsart}
\usepackage{amsmath,amssymb,dsfont}
\usepackage{graphicx}
\usepackage[a4paper,  margin=3.7cm]{geometry}

\title[A Curie-Weiss Theory of the Widom-Rowlinson Model]{A Curie-Weiss Theory of the Continuum Widom-Rowlinson Model}

\author{ Yuri  Kozitsky}

\address{Instytut Matematyki, Uniwersytet Marii Curie-Sk{\l}odowskiej, 20-031 Lublin, Poland}
\email{jkozi@hektor.umcs.lublin.pl}

\author{Mykhailo Kozlovskii}

\address{Institute for Condensed Matter Physics of the National Academy of Sciences of Ukraine, 79011 Lviv, Ukraine }

\email{mpk@icmp.lviv.ua}

\thanks{The present research was supported by the European Commission under the project
STREVCOMS PIRSES-2013-612669.}

\begin{document}

%Use showkeys class option if keyword

\maketitle

\begin{abstract}
A version of the continuum Widom-Rowlinson model is introduced and
studied. It is a two-component gas of point particles placed in
$\mathbf{R}^d$ in which like particles do not interact and unlike
particles contained in a given vessel of volume $V$ repel each other
with intensity $a/V$. This model is thermodynamically equivalent to
a one-component gas with multi-particle interaction. For both
models, a rigorous theory of a phase transition is presented and the
ways of its construction in the framework of the grand canonical
formalism are outlined.

\end{abstract}

%\pacs{64.60.Bd; 64.60.De}% PACS, the Physics and Astronomy
                             % Classification Scheme.

\section{\label{sec1}Introduction}

The rigorous theory of phase transitions in continuum particle
systems has got much more modest results as compared to its
counterpart dealing with lattices, graphs, etc. In fact, there exist
only few models with local interactions in which the existence of a
liquid-vapor phase transition was rigorously proved. One of them is
the model introduced in \cite{WR} by B. Widom and J. S. Rowlinson in
which the potential energy of $n$ point particles located at $x_1,
\dots, x_n \in \mathbf{R}^d$ is set to be $\theta[ W(x_1 , \dots ,
x_n) - n]$, where $\theta>0$ is a parameter and $W$ is the volume of
the area $\cup_{i=1}^n B(x_i)$ covered by the balls of unite volume
centered at the corresponding particles. This model is
thermodynamically equivalent to a two-component system with binary
interactions in which the interaction between unlike particles is a
hard-core repulsion and is zero otherwise. In \cite{R}, D. Ruelle
proved that this model undergoes a phase transition of first order
whenever $d\geq 2$. Later on, this result was extended in
\cite{Kot}, see also \cite{GHM} for a review. However, these
theories give not too much for understanding the details of the
phenomenon. No rigorous results are available on the behavior at the
phase-transition threshold. The very existence of such a threshold
remains unknown for this model. At the same time, for a number of
lattice models the mean field results essentially improve
understanding phase transitions in the corresponding models with
local interactions \cite{Biskup}. It is then quite natural to
develop the mean field theory of phase transitions also in continuum
systems. For the Widom-Rowlinson model, the first attempt to do this
was undertaken already in \cite[Sect. VII]{WR}. Assuming that the
particles are distributed in a given vessel ``at random" the authors
heuristically deduced an equation of state (eq. (7.4) in \cite{WR})
which manifests a first order phase transition. Unfortunately, such
and similar heuristic mean-field theories (sometimes called
\emph{naive}, cf. \cite[page 216]{Simon}) are not free from
mathematical inconsistencies and other drawbacks of mathematical
nature that diminish the value of their results. The first steps in
developing a rigorous mean-field theory of a liquid-vapor phase
transition were made by J. L. Lebowitz and O. Penrose in \cite{LP}
where Kac-type interaction potentials were employed to obtain (in
the canonical formalism) an explicit form of the pressure-density
dependence that describes a first order phase transition. Another
way of developing a rigorous mean field theory is to use Curie-Weiss
interaction potentials, see \cite{EN,E,Zag}. For lattice systems,
this way was formulated as a coherent mathematical theory in the
framework of which thermodynamic phases are constructed as
probability measures on the spin-configuration spaces, see, e.g.,
\cite[Section II]{Ku}. The aim of this letter is to report the
results of a rigorous study of an analog of the Widom-Rowlinson
model with Curie-Weiss interaction potentials made in the grand
canonical formalism.

\section{\label{ssec2.1} The Model}

We deal with infinite systems of point particles placed in the space
$\mathbf{R}^d$. If the particles do not interact, their state of
thermal equilibrium (phase) is a Poisson probability measure $P_z$.
The only parameter characterizing this measure is activity
$z=e^\mu$, where $\mu =$ (physical chemical potential$)/ k_B T$,
$k_B$ and $T>0$ being Boltzmann's constant and absolute temperature,
respectively. For a vessel $\Lambda\subset \mathbf{R}^d$ of finite
volume $V$ and a nonnegative integer $n$, the measure $P_z$ assigns
the probability
\begin{eqnarray}
  \label{1}
P_z(\Gamma_{\Lambda,n}) = \frac{(z V)^n}{ n!} \exp\left( - z
V\right)
\end{eqnarray}
to the event $\Gamma_{\Lambda,n}$: ``$\Lambda$ contains $n$
particles". Let now point particles of two types, 0 and 1, be placed
in the same space $\mathbf{R}^d$. If they do not interact, their
state of thermal equilibrium is the Poisson measure $P_{z_0 ,z_1} =
P_{z_0} \otimes P_{z_1}$ such that  the event $\Gamma_{\Lambda,
n_0}\times \Gamma_{\Lambda, n_1}$ has probability
\begin{equation}
  \label{1a}
P_{z_0,z_1}(\Gamma_{\Lambda,n_0}\times \Gamma_{\Lambda, n_1}) =
P_{z_0}(\Gamma_{\Lambda,n_0}) \cdot P_{z_1}(\Gamma_{\Lambda,n_1}),
\end{equation}
where $ P_{z_i}(\Gamma_{\Lambda,n_i})$, $i=0,1$, are as in (\ref{1})
and the event is ``$\Lambda$ contains $n_0$ particles of type $0$
and $n_1$ particles of type $1$". In this case, the particle
densities are $\varrho_i = z_i = e^{\mu_i}$, $i=0,1$, and the
equation of state reads
\begin{equation*}
 % \label{2}
  p = \varrho_0 + \varrho_1 = e^{\mu_0} + e^{\mu_1},
\end{equation*}
where $p$ is the pressure in the system.

For interacting particles, phases are constructed as limits $\Lambda
\to \mathbf{R}^d$ of local Gibbs measures $P_{z}^{\Lambda,\Phi}$
($P_{z_0,z_1}^{\Lambda,\Phi}$ for two-component systems) describing
the portion of the particles contained in the vessel $\Lambda$ and
interacting with the energy $\Phi$, see, e.g., \cite{GHM}. In this
work,  we introduce two models that -- like the Widom-Rowlinson
model -- can be considered as two versions of the same model. The
first one is a two-component gas of point particles in
$\mathbf{R}^d$. For a vessel $\Lambda \subset \mathbf{R}^d$ of
volume $V$, unlike particles contained in $\Lambda$ repel each other
with intensity $a/V>0$, whereas like particles do not interact.
Thus, the potential energy of the collection of $n_0$ particles of
type 0 located at $x_1^{0}, \dots, x_{n_0}^{0} \in \Lambda$ and of
$n_1$ particles of type 1 located at $x_1^{1}, \dots, x_{n_1}^{1}
\in \Lambda$ is
\begin{equation}
  \label{U1}
 \Phi_V (x_1^{0}, \dots, x_{n_0}^{0}; x_1^{1}, \dots, x_{n_1}^{1}) =
 \sum_{i=1}^{n_0} \sum_{j=1}^{n_1}\frac{a}{V} = \frac{a}{V}n_0 n_1.
\end{equation}
The grand canonical partition function of this collection then is
\begin{eqnarray}
  \label{3}
& & \Xi_V (a, \mu_0 , \mu_1)\\[.2cm] \nonumber &  = & \sum_{n_0, n_1=0}^\infty
 \frac{1}{n_0! n_1 !} \int_{\Lambda^{n_0}}
\int_{\Lambda^{n_1}} \exp\left( \mu_0 n_0 + \mu_1 n_1 -
\frac{a}{V}n_0 n_1\right)d x_1^{0} \cdots d
x_{n_0}^{0} d x_1^{1} \cdots d x_{n_1}^{1}  \\[.2cm]\nonumber &= & \sum_{n_0, n_1=0}^\infty
 \frac{V^{n_0 + n_1}}{n_0! n_1 !}  \exp\left(
\mu_0 n_0 + \mu_1 n_1 - \frac{a}{V}n_0 n_1\right).
\end{eqnarray}
Here the interaction parameter $a>0$ and the chemical potentials
$\mu_i\in \mathbf{R}$, $i=1,2$, include the reciprocal temperature
$\beta$ and thus are dimensionless. The second our model is a
one-component system of point particles interacting as follows. For
a vessel $\Lambda$ of volume $V$, the potential energy of the
collection of $n$ particles located at $x_0, \dots , x_n\in \Lambda$
is set to be
\begin{equation*}
 % \label{2a}
\widehat{\Phi}_V (x_1 , \dots , x_n) = V \theta \left[ 1 -
\exp\left( - \frac{a}{V}
 n\right)\right].
\end{equation*}
Here $\theta>0$ is a parameter, similar to that in \cite{WR}
mentioned above. Then the corresponding grand canonical partition
function is
\begin{eqnarray}
  \label{4a}
 \widehat{\Xi}_V (a, \mu , \theta) &  = & \sum_{n=0}^\infty
 \frac{1}{n!} \int_{\Lambda^{n}} \exp\left( \mu n - V \theta \left[ 1 -
\exp\left( - \frac{a}{V}
 n\right)\right]
\right)d x_1 \cdots d
x_{n} \\[.2cm]\nonumber &= & \sum_{n=0}^\infty
 \frac{V^{n}}{n!}  \exp\left(
\mu n  - V \theta \left[ 1 - \exp\left( - \frac{a}{V}
 n\right)\right] \right) \\[.2cm]\nonumber &= & \exp\left(
- V\theta\right) \Xi_{V}(a,\mu,\ln \theta).
\end{eqnarray}
The latter equality can readily be derived by summing out in
(\ref{3}) over $n_1$. The dependence of the pressure $p$ in the
two-component system (resp. $\widehat{p}$ in the one-component
system) on $a$ and $\mu_i$, $i=0,1$ (resp. on $a$, $\theta$ and
$\mu$) is then obtained in the thermodynamic limit
\begin{eqnarray}
  \label{4}
  p = p(a,\mu_0, \mu_1)& = & \lim_{V\to +\infty} \frac{1}{V} \ln
  \Xi_{V}(a,\mu_0,\mu_1),\\[.2cm] \nonumber
  \widehat{p} =  \widehat{p}(a,  \theta, \mu)& = & \lim_{V\to +\infty} \frac{1}{V} \ln
   \widehat{\Xi}_{V}(a,\theta,\mu),
\end{eqnarray}
which by (\ref{4a}) yields $\widehat{p}= p - \theta$. Thus, the
particle density $\varrho$ in the one-component system and the
density $\varrho_0$ of the particles of type 0 in the two-component
system are related to each other by
\begin{equation}
  \label{4x}
\varrho = \frac{\partial \widehat{p}}{\partial \mu} = \frac{\partial
p}{\partial \mu_0}\bigg{|}_{\mu_0=\mu, \ \mu_1 = \ln \theta} =
\varrho_0.
\end{equation}
Therefore, the study of the one-component system amounts to studying
its two-component counterpart by employing (\ref{3}) and (\ref{4x}).

\section{\label{ssec2.2} The Results}

In view of (\ref{3}), we will deal with three thermodynamic
variables $a, \mu_0, \mu_1$, and hence with the phase space
$$\mathcal{F}= \{ (a, \mu_0, \mu_1): a \geq 0, \mu_0 , \mu_1 \in
\mathbf{R}\}.$$ Our main result is the statement that $\mathcal{F}$
can be divided into three disjoint subsets, i.e., presented as
\begin{equation}
  \label{4aa}
  \mathcal{F}= \mathcal{R}\cup \mathcal{C} \cup \mathcal{M}.
\end{equation}
A point $(a,\mu_0, \mu_1)$ belongs to one of these subsets according
to the number of global maxima of the function
\begin{equation}
  \label{10}
  E(y) = f(a,\mu_0 +y) + f(a,\mu_1 - y) - \frac{y^2}{2a},
\end{equation}
where
\begin{equation}
  \label{11}
  f(a,x)  =  \frac{a}{2} \left[ u(a,x)\right]^2 + u(a,x), \quad x\in
  \mathbf{R}.
\end{equation}
Here $u$ is a special function, which can be expressed through
Lambert's $W$-function\cite{W}  as follows
\begin{equation}
  \label{8}
  u(a,x) = \frac{1}{a} W ( a e^x), \qquad x \in \mathbf{R}.
\end{equation}
For a fixed $a>0$, the function $\mathbf{R}\ni x \mapsto u(a, x)$
can be obtained as the inverse to $(0, +\infty)\ni u \mapsto x(u) =
a u + \ln u$, by which one gets that
\begin{eqnarray}
  \label{8a}
& & u(a,x) \exp\left[a u(a,x)  \right] = e^x,\\[.2cm] \nonumber
  & & \frac{\partial }{\partial x} u(a, x)  =  \frac{ u(a, x)}{1 + a u(a,
  x)}.
\end{eqnarray}
By (\ref{10}) and (\ref{8a}) it follows that the points of global
maximum of $E$ are also its local maximum points and hence can be
obtained from the equation
\begin{equation}
  \label{12}
 y = a u(a, \mu_0 + y ) - a u(a, \mu_1-y), \quad y\in \mathbf{R}.
\end{equation}
The single-phase domain $\mathcal{R}= \mathcal{F}\setminus
(\mathcal{C}\cup \mathcal{M})$, cf. (\ref{4aa}), consists of all
those $(a, \mu_0, \mu_1)$ for each of which there exists a unique
global maximum of $E$ (at some $y_*$). Then the unique phase
existing at this point $(a, \mu_0, \mu_1)$  is the Poisson state
$P_{\tilde{z}_0 , \tilde{z}_1}$, see (\ref{1}) and (\ref{1a}), where
\begin{equation}
  \label{7}
\tilde{z}_0 = u(a,\mu_0 + y_*) , \quad \tilde{z}_1 = u(a,\mu_1 -
y_*).
\end{equation}
The set
\begin{equation}
  \label{5}
  \mathcal{M} := \{ (a, \mu, \mu) : a>0, \mu > 1 - \ln a\}
\end{equation}
consists of phase coexistence points, and
\begin{equation}
  \label{6}
  \mathcal{C}:=\{ ( a , 1- \ln a , 1 - \ln a): a >0\}
\end{equation}
is the line of critical points. For $(a,\mu_0, \mu_1) \in
\mathcal{M}$ (i.e., for $\mu_0 = \mu_1 = \mu > 1- \ln a$), the
function in (\ref{10}) has two equal maxima located at $\pm
\bar{y}(a,\mu)$, where the order parameter $\bar{y}(a,\mu)>0$ is the
unique solution of the equation $\psi (y ) = \mu - (1-\ln a)$ with
\begin{equation*}
 % \label{6b}
\psi (y ) = y + \frac{y}{e^y -1} - 1 + \ln \frac{y}{e^y -1}, \quad
y>0 .
\end{equation*}
For small $y>0$, we have that $\psi(y) = y^2/24 + o(y^2)$. Thus,
$$\bar{y}(a, \mu) = \sqrt{24(\mu-(1-\ln a))} + o(\mu-(1-\ln a))$$ for
small positive $\mu - (1-\ln a)$. For $\bar{y}(a, \mu)>0$, there
exist two phases $P_{\tilde{z}^{+} , \tilde{z}^{-}}$ and
$P_{\tilde{z}^{-} , \tilde{z}^{+}}$, where
\begin{equation*}
  %\label{9}
\tilde{z}^{\pm} = u(a, \mu \pm \bar{y}(a,\mu)).
\end{equation*}
The equation of state of the two-component system has the following
form, cf. (\ref{4}),
\begin{equation}
  \label{9a}
 p = a \varrho_0 \varrho_1 + \varrho_0 + \varrho_1,
\end{equation}
where the densities $\varrho_i$ are, cf.  (\ref{7}),
\begin{equation}
  \label{9b}
  \varrho_0 = u(a, \mu_0 + y_*), \quad \varrho_1 = u(a, \mu_1 -
  y_*).
\end{equation}
Note that each $\varrho_i$ depends on both $\mu_0$, $\mu_1$ and
$\varrho_i = \frac{\partial p}{\partial \mu_i}$, $ i=0,1$, cf.
(\ref{4x}). Note also that the densities satisfy
\begin{equation}
  \label{9d}
\varrho_0 = \exp\left(\mu_0 - a \varrho_1 \right), \quad \varrho_1 =
\exp\left(\mu_1 - a \varrho_0 \right).
\end{equation}
That is, due to the repulsion both densities are smaller than in the
free case $a=0$.

Let us turn now to the ground states of the two-component model
which one obtains by passing to the limit $a\to +\infty$. To this
end, we consider $\varrho_i$, $i=0,1$, as differentiable functions
of $a$ defined in (\ref{9d}). Let $\dot{\varrho}_i$, $i=0,1$, stand
for the corresponding $a$-derivatives. Differentiating both sides of
each equality in (\ref{9d}) after some calculations we get
\begin{equation}
  \label{9e}
\dot{\varrho}_0 - \dot{\varrho}_1 = \frac{a \varrho_0 \varrho_1}{1 -
a^2 \varrho_0 \varrho_1}\left( \varrho_0 - \varrho_1 \right).
\end{equation}
The denominator here is positive by the fact that $y_*$ used in
(\ref{9b}) is the point of local maximum of $E$ given in (\ref{10}).
For $\mu_0
> \mu_1$, we have that $\varrho_0
> \varrho_1$ for all $a>0$. Indeed, assuming
$\varrho_0= \varrho_1$ for some $a>0$, we then get by (\ref{9d})
that $e^{\mu_0}=e^{\mu_1}$, which contradicts the assumed inequality
$\mu_0
> \mu_1$.
Thus, by (\ref{9e}) $\varrho_0 - \varrho_1$ is an increasing
function of $a$, which yields $\varrho_0 - \varrho_1 \geq \varrho_0
- \varrho_1|_{a=0} = e^{\mu_0} - e^{\mu_1}$. By (\ref{9d}) and the
latter estimate we then get
\begin{eqnarray}
  \label{9f}
 \varrho_1 & = & \varrho_0 \exp\left(- (\mu_0 -\mu_1) - a (\varrho_0 - \varrho_1)
 \right) \\[.2cm] \nonumber
& \leq & \varrho_0 \exp\left(- (\mu_0 -\mu_1) - a (e^{\mu_0} -
e^{\mu_1})
 \right).
\end{eqnarray}
Since $\varrho_0 \leq e^{\mu_0}$, see (\ref{9d}), by (\ref{9f}) we
get that $a\varrho_1 \to 0$, and hence $\varrho_1 \to 0$, as $a\to
+\infty$. At the  same time, $\varrho_0 \geq \varrho_1 +
(e^{\mu_0}-e^{\mu_1})$, which by (\ref{9d}) yields that $\varrho_0
\to e^{\mu_0}$ as $a\to +\infty$. By (\ref{7}) and (\ref{9b}) we
thus conclude that the two-component system has two ground states:
$P_{z_0,0}$ and $P_{0,z_1}$. In each of them, there is only one free
component.

Turn now to the one-component system. Its equation of state reads
\begin{equation}
  \label{9g}
 \widehat{p} = a \theta \varrho e^{- a \varrho} + \varrho - \theta
 \left(1 - e^{- a \varrho} \right),
\end{equation}
that can be obtained by (\ref{9d}) and the formula $\widehat{p}= p-
\theta$. Here $\varrho$ is a function of $\mu\in \mathbf{R}$
obtained from (\ref{9b}), i.e., $\varrho= u(a, \mu+y_*)$. It is
increasing and continuous whenever $\theta \leq e/a$. For $\theta
> e/a$, $\varrho$ makes a jump at $\mu=\ln \theta$ with one-sided
limits $\lim_{\mu \to \ln \theta \pm 0}\varrho = u(a, \mu\pm
\bar{y}(a,\mu))$. That is, the system undergoes a first-order phase
transition with the increment of the density $\varDelta \varrho =
\bar{y}(a,\mu)/a$. The expression in (\ref{9g}) can also be used to
define $\widehat{p}$ as a function of $\varrho$. Namely,
$\widehat{p}$ is as in (\ref{9g}) for $\varrho \leq \tilde{z}^{-}_0$
and $\varrho \geq \tilde{z}^{+}_0$, and $\widehat{p}\equiv
\widehat{p}_* := a \tilde{z}^{+}_0 \tilde{z}^{-}_0 +
\tilde{z}^{+}_0+ \tilde{z}^{-}_0 - \theta$ for
$\varrho\in[\tilde{z}^{-}_0 , \tilde{z}^{+}_0]$.   Note that the
equation of state in (\ref{9g}) with $a=1$ formally coincides with
that found heuristically in \cite{WR}, in which, however, the
horizontal part $\widehat{p}\equiv \widehat{p}_*$ should be found
from the Maxwell rule, see \cite{KK} for more detail.

\begin{figure}[h!]
\includegraphics[width=230pt]{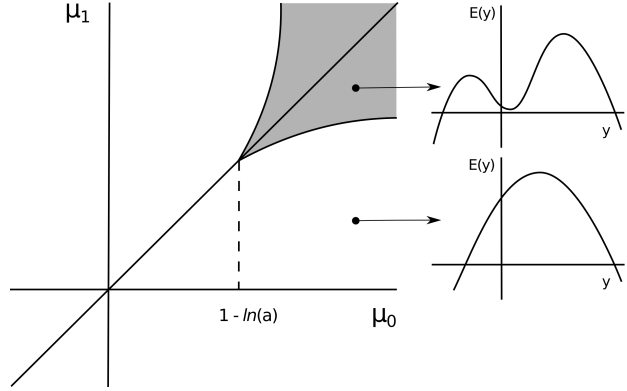}
\caption{Phase diagram at fixed $a$ }\label{F1}
\end{figure}

The part of the phase diagram of the two-component system in the
plane in $\mathcal{F}$ with constant $a>0$ is presented in Fig
\ref{F1}.
Points from the grayed area correspond to the existence of three
solutions of (\ref{12}), one of which is $y_*$. Note that $y_* >0$
for $\mu_0 > \mu_1$. At the boundaries of this area (symmetric under
 $\mu_0 \leftrightarrow \mu_1$), (\ref{12}) has only two
solutions. The upper branch of the boundary is described by the
equation $$\eta = \sqrt{\xi^2 -1} + \ln \left(\xi - \sqrt{\xi^2 -1}
\right),$$ where $\xi = (\mu_0 + \mu_1 )/2 + \ln a$ and $\eta =
(\mu_1 - \mu_0)/2$. For all points from the complement to the grayed
area, (\ref{12}) has only one solution. Note that $y_* =0$ for
$\mu_0 = \mu_1 < 1 - \ln a$.

%\subsubsection{\label{sssec1} The order parameter and the equation of state}

\section{Deriving the Results}

Here we outline the way of deriving the results presented in the
preceding section. In \cite{KK}, these results are formulated and
proved as mathematical statements.

By means of the identity $$- \frac{a}{V}n_0 n_1 = -
\frac{a}{2V}n^2_0 - \frac{a}{2V}n^2_1 + \frac{a}{2V}(n_0 - n_1)^2,$$
and by a standard Gaussian formula we rewrite (\ref{3}) as
\begin{equation}
  \label{20}
\Xi_V (a, \mu_0 , \mu_1) = \sqrt{\frac{V}{2\pi a}}
\int_{-\infty}^{+\infty} \exp\left(V E_V (y) \right) d y,
\end{equation}
with
\begin{equation*}
  %\label{21}
 E_V(y) = f_V (a, \mu_0 +y) + f_V(a, \mu_1-y) - \frac{y^2}{2a}.
\end{equation*}
Here $f_V$ is defined by the following relation
\begin{equation}
  \label{22}
\exp\left( V f_V (a, x)\right) = \sum_{n=0}^\infty \frac{V^n}{n!}
\exp\left( x n - \frac{a}{2V} n^2 \right),
\end{equation}
and thus is an infinitely differentiable function of $x\in
\mathbf{R}$ for each fixed $a>0$ and $V>0$. Then so is $ E_V$ as a
function of $y\in \mathbf{R}$. In view of (\ref{4}) we have to find
the large $V$ asymptotic in (\ref{20}). To this end we employ a more
advanced version of Laplace's method as $E_V$ depends on $V$.
According to \cite[Theorem 2.2]{Fed}, this amounts to proving that
the third $x$-derivative of $f_V$ is bounded as $V\to +\infty$,
which is the most challenging aspect of the theory. Taking the
$x$-derivative of both sides of (\ref{22}) we obtain that $u_V (a,
x) := \partial f_V(a,x) /
\partial x$ satisfies, cf. (\ref{8a}),
\begin{eqnarray*}
  %\label{23}
u_V (a,x) \exp\left( \int_0^1 u_V \left( a, x- \frac{a}{V} t\right)
dt\right)= \exp\left( x -  \frac{a}{2V} \right).
\end{eqnarray*}
On the other hand, also by (\ref{22}) we get that
\begin{equation*}
  %\label{23}
u_V (a,x)  = \langle n \rangle /V = \frac{1}{V} \sum_{n=1}^\infty n
\pi_n,
\end{equation*}
with $$\pi_n =   \frac{V^n}{n!} \exp\left( n x - \frac{an^2}{2V}
 \right) \bigg{/} \sum_{n=0}^\infty  \frac{V^n}{n!} \exp\left( n x - \frac{an^2}{2V}
 \right).$$
Then the consecutive $x$-derivatives of $u_V$ are $$u'_V (a, x) =
\frac{1}{V} \langle (n-\langle n\rangle)^2 \rangle, \ \ u''_V (a, x)
= \frac{1}{V} \langle (n-\langle n\rangle)^3 \rangle.$$ These
formulas allow for obtaining uniform in $V$ upper bounds for $u'_V
(a, x)$ and $|u''_V (a, x)|$. Thereafter, we apply the mentioned
version of Laplace's method by which the problem of calculating $p$
in (\ref{4}) is reduced to finding the global maxima of $E$ given in
(\ref{10}). This yields, in particular, that $p$ is given by
(\ref{9a}). We also prove that $f_V \to f$, $u_V \to u$ and $u'_V
\to u'$ as $V\to \infty$, where $f$, $u$ and $u'$ are as in
(\ref{11}), (\ref{8}) and (\ref{8a}), respectively. The proof that
the limiting states are $P_{\tilde{z}_0, \tilde{z}_1}$ with
$\tilde{z}_i$ given in (\ref{7}) is done by showing that the
correlation functions of the local Gibbs measures, see
\cite{Ruelle}, converge as $V\to +\infty$ to those of
$P_{\tilde{z}_0, \tilde{z}_1}$. Here we also employ the result of
\cite{Fed} mentioned above, as well as the convergence of the
derivatives of $f_V$ just mentioned. Then the representation of
$\mathcal{F}$ in (\ref{4a}) with $\mathcal{M}$ and $\mathcal{C}$ as
in (\ref{5}) and (\ref{6}), respectively, are obtained by studying
the global maxima of $E$.

\section{\label{ss3} Concluding Remarks}

We recall that the parameters $a$, $V$, $\mu_0$ and $\mu_1$ in the
expression for $\Xi_V$ in (\ref{3}) are dimensionless. By chosen
them we, in fact, fix some metric of the habitat space
$\mathbf{R}^d$. The change of this metric (scale) can be done by
passing to $V^\alpha := \alpha V$ for some scale parameter
$\alpha>0$. This, of course, leads to the change $\varrho_i \to
\varrho_i^\alpha = \varrho_i/\alpha$, $i=0,1$, and also to $a\to
a^\alpha = \alpha a$ and $p\to p^\alpha = p/\alpha$, cf. (\ref{U1})
and (\ref{9a}), respectively. Then by (\ref{9d}) we obtain that the
rescaled densities and the interaction parameter $a^\alpha$ satisfy
\[
\varrho_0^\alpha = \exp\left( \mu_0^\alpha - a^\alpha
\varrho_1^\alpha\right), \qquad \varrho_1^\alpha = \exp\left(
\mu_1^\alpha - a^\alpha \varrho_0^\alpha\right),
\]
with $\mu_i^\alpha = \mu_i - \ln \alpha$, $i=0,1$. By this we
conclude that the description of the phases corresponding to the
points $(a, \mu_0, \mu_1)\in \mathcal{F}$ is scale-invariant. Thus,
one can choose $\alpha = 1/a$ and hence consider the plane in
$\mathcal{F}$ with $a=1$. Then the results corresponding to a
general point $(a, \mu_0, \mu_1)\in \mathcal{F}$ can be obtained
from those obtained for $a=1$ by the rescaling as just described.
Note, however, that in view of taking the thermodynamic limit $V\to
+\infty$, the role of the scale of $V$ may get be lost. Then
considering $a$ as an interaction parameter helps to reveal it.
Moreover, from our analysis it follows that the phase transition
occurs for an arbitrarily small interaction $a$. Also for the
following reasons it might be worth to deal with general values of
$a$: (i) to be able to pass to the free case $a=0$; (ii) to obtain
the ground states in the limit $a\to +\infty$ as described above;
(iii) to get clues on the phase transition in the original
Widom-Rowlinson model in which the metric is rather fixed by the
choosing the radius of the hard-core repulsion. In this model the
interaction energy of the collection of $n_0$ particles of type zero
located at $x_1^0, \dots , x_{n_0}^0$ with $n_1$ particles of type
one located at $x_1^1, \dots , x_{n_1}^1$ is written in the form,
cf. \cite[Section 10.2]{GHM},
\begin{equation}
  \label{WRo}
\Phi(x_1^0, \dots , x_{n_0}^0; x_1^1, \dots , x_{n_1}^1) =
\sum_{k=1}^{n_0} \sum_{l=1}^{n_1} \infty \mathbf{1}_{|x_k^0 -
x_l^{1}|\leq r_d}(x_k^{0}, x_l^{1}),
\end{equation}
where $\mathbf{1}_{|x_k^0 - x_l^{1}|\leq r_d}$ is the corresponding
characteristic function and $r_d$ is the radius of the ball in
$\mathbf{R}^d$ of unit volume. In contrast to our $\Phi_V$ given in
(\ref{U1}) $\Phi$ takes values either zero or infinity, and hence no
parameter like our $a$ can be associated with this model. Regarding
its thermodynamic phases the following is known. There exist $\mu_*,
\mu^* \in \mathbf{R}$ such that $\mu_*<\mu^*$ and: (a) for $\mu_0
=\mu_1 =\mu<\mu_*$, there exists only one thermodynamic phase; (b)
for $\mu_0 =\mu_1 =\mu>\mu^*$, there exist at least two
thermodynamic phases. In contrast to our description, nothing is
known about the threshold that might exist on the interval $[\mu_*,
\mu^*]$. Having in mind that for spin systems on $\mathbf{Z}^d$ with
local (properly normed) interactions the mean-field theory of a
phase transition becomes exact in the limit $d\to +\infty$, see
\cite[Theorem II.14.1, page 228]{Simon}, one might speculate that
also for the model in (\ref{WRo}) there exists a threshold value
$\mu_c = 1 - \ln a_{\rm WR} (d)\in [\mu_*, \mu^*]$ with $a_{\rm
WR}(d)$ satisfying $a_{\rm WR}(d) \to a_{\rm WR}$ as $d\to +\infty$
for some $a_{\rm WR}>0$. If this is true, then our model with a
particular value $a=a_{\rm WR}$ can be considered as a mean-field
limit of the original Widom-Rowlinson model.

%\bibliography{kokobib}% Produces the bibliography via BibTeX.

\end{document}